\documentclass[twocolumn]{aastex631}

\shorttitle{Double Tilted Rowland Spectrograph}
\shortauthors{G\"unther et al.}
\graphicspath{{./}{figures/}{ifigs/}}

\usepackage{amsmath}

\begin{document}

\title{Concept of a Double Tilted Rowland Spectrograph for X-rays}


\correspondingauthor{Hans Moritz G\"unther}
\email{hgunther@mit.edu}

\author[0000-0003-4243-2840]{Hans Moritz G{\"u}nther}
\affiliation{MIT Kavli Institute for Astrophysics and Space Research, 77 Massachusetts Avenue, Cambridge, MA 02139, USA}
\author[0000-0002-9184-4561]{Casey T. DeRoo}
\affiliation{Dept. of Physics \& Astronomy, University of Iowa, Iowa City, IA 52242, USA}
\author[0000-0001-9980-5295]{Ralf K. Heilmann}
\affiliation{MIT Kavli Institute for Astrophysics and Space Research, 77 Massachusetts Avenue, Cambridge, MA 02139, USA}
\affil{Space Nanotechnology Laboratory, MIT Kavli Institute for Astrophysics and Space Research, Cambridge, MA 02139, USA}
\author[0000-0002-6747-9648]{Edward Hertz}
\affiliation{Center for Astrophysics, Harvard-Smithsonian Astrophysical Observatory, Cambridge, MA 02138, USA}

\begin{abstract}
High-resolution spectroscopy in soft X-rays ($<2$~keV) requires diffractive elements to resolve any astrophysically relevant diagnostics, such as closely spaced lines, weak absorption lines, or line profiles. The Rowland torus geometry describes how gratings and detectors need to be positioned to optimize the spectral resolving power. We describe how an on-axis Rowland geometry can be tilted to accommodate blazed gratings. In this geometry, two channels with separate optical axes can share the same detectors (double tilted Rowland spectrograph, DTRS). Small offsets between the channels can mitigate the effect of chip gaps and reduce the alignment requirements during the construction of the instrument. The DTRS concept is especially useful for sub-apertured mirrors, because it allows an effective use of space in the entrance aperture of a spacecraft.
One mission that applies this concept is the Arcus Probe.

\end{abstract}


\section{Introduction} \label{sec:intro}
X-ray spectroscopy can deliver many insights into the astrophysics of the hot and energetic universe that are not otherwise observable. In particular, high-resolution spectroscopy (with resolving power $R=E/\Delta E> 500$ or better) can resolve closely-spaced emission lines which offer diagnostics such as density and temperature, reveal the near-edge fine structure in absorption lines, or allow us to detect the presence of faint and narrow absorption lines caused by the warm-hot intergalactic medium in front of bright background quasars to give just a few examples. In the X-ray band, such high resolving powers can be achieved with either microcalorimaters or dispersive elements such as crystals or diffraction gratings. Microcalorimeters offer a fixed absolute energy resolution $\Delta E$, which gives them good resolving power $R$ at large energies $E$ and consequently a microcalorimeter has been launched on Astro-E/Hitomi/SXS \citep{2014SPIE.9144E..2AM} and XRISM/Resolve \citep{2018JATIS...4a1214K} and this technology is also considered for other large missions in development such as Athena \citep{2014SPIE.9144E..2LR} and Lynx \citep{2019JATIS...5b1017B}. Below about 2~keV diffraction gratings have a distinct advantage.

The two largest X-ray observatories in history both field diffraction gratings \citep[see][for a review]{2010SSRv..157...15P}: The Reflection Grating Spectrometer (RGS) on XMM-Newton \citep{2001A&A...365L...7D} and both the High-Energy Transmission Grating  \citep[HETG,][]{2005PASP..117.1144C} and the Low-Energy Transmission Grating \citep[LETG,][]{1997SPIE.3113..172P} on Chandra.
Unlike microcalorimeters, grating spectrographs require an additional dispersive element (a transmission or a reflection grating) in the optical path of the instrument and dedicated detectors to capture the dispersed signal. For a given diffraction order, gratings typically deliver a fixed wavelength resolution $\Delta \lambda$, so they perform better than microcalorimeters at longer wavelengths/lower energies. One common layout for a diffraction spectrometer is the Rowland torus \citep{Beuermann:78}. In the plane defined by the optical axis and the dispersion direction, the gratings are located at one side of a circle, the detectors on the other. This circle is commonly called the ``Rowland circle''. In three dimensions, the surface that defines the best grating positions has the shape of a torus, thus the name ``Rowland torus'' for this geometry (see Sect.~\ref{sect:onetorus} for a figure).

In this article, we present an optical layout for an instrument that stacks two or four different channels with distinct optical axes in such a way that each channel follows the Rowland torus prescription, yet at the same time all channels are imaged onto a common set of detectors.

This concept has been developed and described in the context of the Arcus mission, which was originally proposed as a NASA Medium Explorer \citep{2019SPIE11118E..0WS} and  has since been proposed to the NASA Astrophysics Probe call \citep{2023SPIE12678E..0ES}. However, the DTRS concept is applicable to mission architectures beyond that of Arcus, and could be applied to a broad range of focal lengths, grating parameters, etc. This paper aims to discuss the advantages and challenges of this layout from a general perspective. We note that this discussion will be necessarily qualitative (rather than quantitative) as performance metrics such as resolving power and effective area depend on the design parameters that will be specific to a given instrument (e.g., focal length, efficiency).

The structure of this article is as follows: First, we discuss the properties of the optical elements that are most influential for the design of a DTRS, namely mirror properties, gratings, and detectors in Sect.~\ref{sect:elements}. Then, we illustrate the optical layout, starting by modifying the well-known single-channel Rowland torus spectrograph design step-by-step to a DTRS. In section~\ref{sect:discussion}, we discuss advantages and disadvantages of our DTRS concept in general and for the example of Arcus in particular (section~\ref{sect:applications}). We end with a short summary in section \ref{sect:summary}.

\section{Elements of an X-ray grating spectrograph}
\label{sect:elements}
An X-ray spectrograph using the Rowland layout consists of a focusing mirror, diffractive gratings, and a detector.

\subsection{Mirror characteristics and sub-aperturing}
\begin{figure*}
\plotone{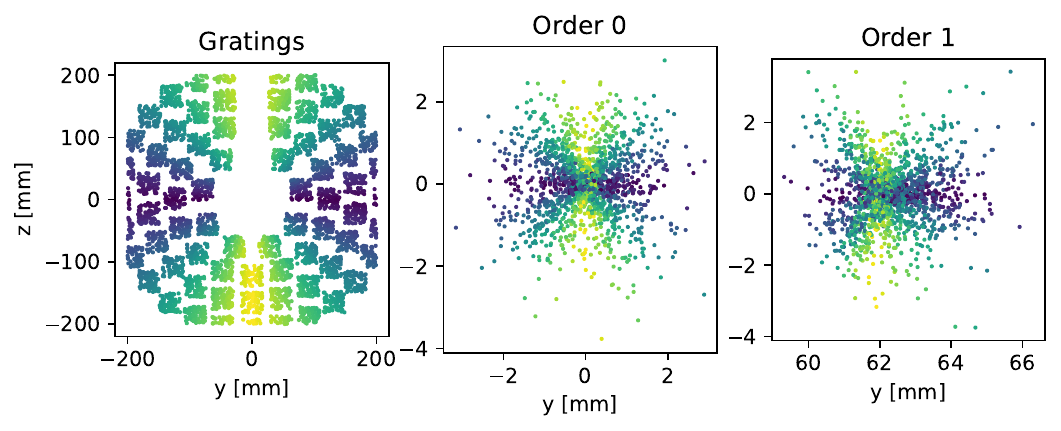}
\caption{
    Illustration of the concept of sub-aperturing. All panels show a view from the universe looking down onto the telescope. \emph{left}: Each dot is a photon shown as it passes through a diffraction grating. Photons are colored according to their polar angle in this panel. \emph{middle and right:} Photons detected in the zeroth and first diffraction order on the detector, respectively. The dispersion direction is from left to right ($y$ coordinate), the distance from the focal point at $(y, z) = (0, 0)$ is shown. Each photon is shown in the same color as in the left panel. Photons passing mirror and grating close to $z=0$ (dark purple) are spread out along the dispersion direction ($y$ coordinate); photons passing mirror and gratings close to $y=0$ are more concentrated along the dispersion direction. Sub-aperturing means that the mirror and gratings do not cover a full circle, but only regions marked in yellow and green in the left panel, so that the resulting photon distributions on the detector are narrower in dispersion direction, which increases the spectral resolving power.}
\label{fig:subaperture}
\end{figure*}

The concept of a DTRS can be implemented with different mirror technologies. In X-rays, the mirror assembly will typically consist of nested (full or partial) shells that reflect X-rays in grazing incidence. The higher the energy of the X-ray, the smaller the angle between the mirror surface and the incoming ray has to be to allow for efficient reflection. The size of the collecting area is to first order proportional to the number of photons that the spectrograph can detect, and the spectral resolving power $R=\lambda / \Delta \lambda$ increases when the mirror point-spread-function (PSF) decreases. Depending on the type of mirror chosen, the PSF can be dominated by the alignment error between mirror shells, in particular for mirror assemblies that consist of hundreds of elements. In this case, X-rays are offset from the focus point both in the plane of reflection and perpendicular to it.
On the other hand, mirrors can also have a figure error, which is the deviation of the mirror surface from the ideal shape. If the PSF is dominated by the figure error or scattering by microroughness or particulates, then scattering angles in the plane of reflection are typically larger than the out-of-plane scatter. In this case, sampling a limited azimuthal span of the aperture -- subaperturing -- can improve the spectral resolving power \citep{1987ApOpt..26.2915C,2010SPIE.7732E..1JH}. This is illustrated in Fig.~\ref{fig:subaperture}.

\subsection{Dispersive elements}
In a Rowland torus geometry, diffraction gratings are placed on the surface of the Rowland torus, ideally tangentially for transmission gratings. They diffract light according to the grating equation for photons arriving perpendicular to the grating surface:
\begin{equation}
n \lambda = d \sin \theta \label{eqn:diffraction}
\end{equation}
where $\lambda$ is the wavelength of the light, $n$ is the order of the diffraction, $d$ is the grating constant, and $\theta$ is the angle between the outgoing light and the normal to the grating. (The equation can be rewritten for rays hitting the grating at arbitrary angles, but for the purpose of this conceptual discussion this is not needed.)
Typically, these transmission gratings are flat and only a few cm$^2$ in size; the spacing between gratings bars $d$ is the same for all gratings.
While a flat grating necessarily deviates from the curved surface of a torus, in today's instruments those deviations are small enough that they do not impact the resolving power of the spectrograph significantly. Larger gratings would increase the effective area of the instrument because less space is needed for frames and other mounting structures that hold the gratings in place, but they would also deviate more from the surface of the torus thus reducing $R$.
To prevent a loss of $R$ with larger gratings one can use gratings where $d$ varies across the grating \citep{1986SPIE..560...96H,2020SPIE11444E..88G} or bend the gratings \citep{2020SPIE11444E..88G}. \citet{2019SPIE11118E..11G} shows that gratings can be bent to the required radii without degrading their performance.

\subsection{Detectors}
In our DTRS design several spectral traces are positioned in parallel (section~\ref{sect:opticallayout}), so the detector has to be position sensitive to separate individual channels.  Because it covers a larger area than in direct imaging observations, a detector with moderate cost and power requirements per surface area is well-suited for this design. The detector should follow the shape of the Rowland circle. This can be achieved by using curved detectors or by tiling smaller flat detector elements. 

The gratings spread out the signal over a large region on the detector. Thus, the count rates per pixel are lower than they would be in an imaging observation. Detectors that might be at risk of pile-up in direct observations can be used for the dispersed signal. On the other hand, the larger regions needed to extract the signal imply that background events (internal background and diffuse astrophysical emission) make up a larger fraction of the extracted signal than in direct imaging observations.
Intrinsic energy resolution is required to both suppress the background by filtering out events that are incompatible with wavelengths of the dispersed signal at any particular location, and to separate grating orders where photons of different orders are diffracted to the same position (e.g.\ the sixth order for the O~{\sc vii} forbidden line of the He-like triplet at 2.21~nm and the seventh order of 1.94~nm photons, close to the O~{\sc viii} Lyman $\alpha$ line end up at almost the same $\theta$ according to equation~\ref{eqn:diffraction}).

In our DTRS design, the images ($n=0$) of all channels are positioned on detectors to provide an accurate wavelength calibration, but also because the high-energy part of the spectrum, which is not efficiently dispersed by gratings, can provide valuable context for the scientific interpretation of the high-resolution grating spectrum. Thus, the detector type chosen should also be able to handle direct imaging, albeit at a lower count rate than in a direct imaging instrument without dispersive elements.

\section{Optical Layout}
\label{sect:opticallayout}
In this section, we describe the optical layout for a DTRS. We start from a single central Rowland torus as in Chandra/HETGS and Chandra/LETGS. We then show how this design can be modified step-by-step.
The general concept is shown using ray-traces with the marxs code \citep{2017AJ....154..243G} using a simple setup for illustration.
For example, the position of the diffractive gratings shown leaves valuable mirror area uncovered and is done without considering the size and location of support structures that the gratings can be mounted on. The grating constant $d$ is chosen to be smaller than feasible for a real instrument.

The points $\vec p$ on the surface of a torus can be parameterized by two angles $\varphi$ and $\theta$.
\begin{equation}
\vec p(\varphi, \theta) = \vec c + R \; \vec e_R(\varphi) + r \; (\vec e_z \sin \theta + \vec e_R(\varphi) \cos \theta)
\end{equation}
where $\vec c$ points to the center of the torus. The axis of symmetry is given by a unit vector $\vec e_z$. We define a vector $\vec e_R(\varphi) = \vec e_x \cos\varphi + \vec e_y \sin\varphi$. $R \vec e_R$ points from the center of the torus to the center of a circle with radius $r$. Sweeping that circle around the symmetry axis forms the torus. If $r < R$, the torus has an inner hole like a doughnut, and if $r > R$ the inner parts of the torus overlap with each other (``spindle torus'').

\begin{figure*}
    \includegraphics[width=0.5\textwidth]{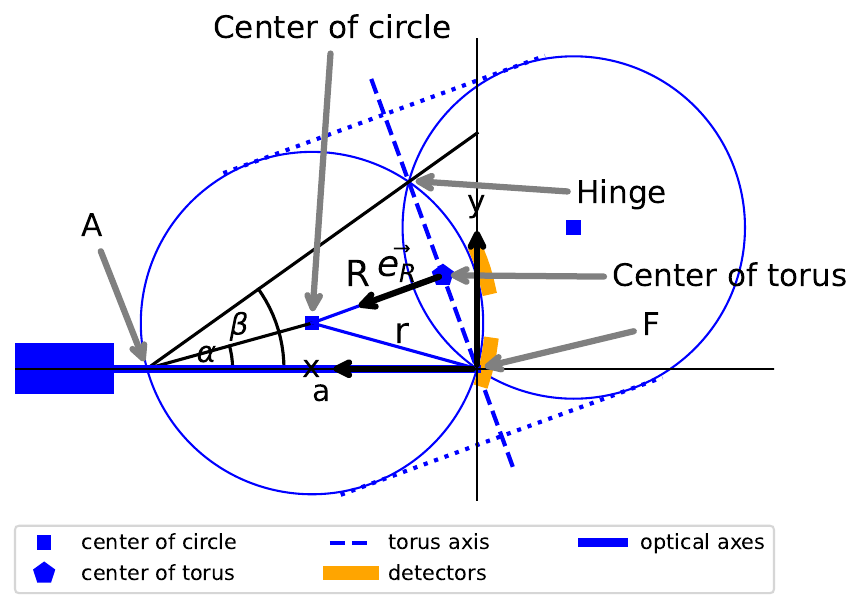}
    \includegraphics[width=0.5\textwidth]{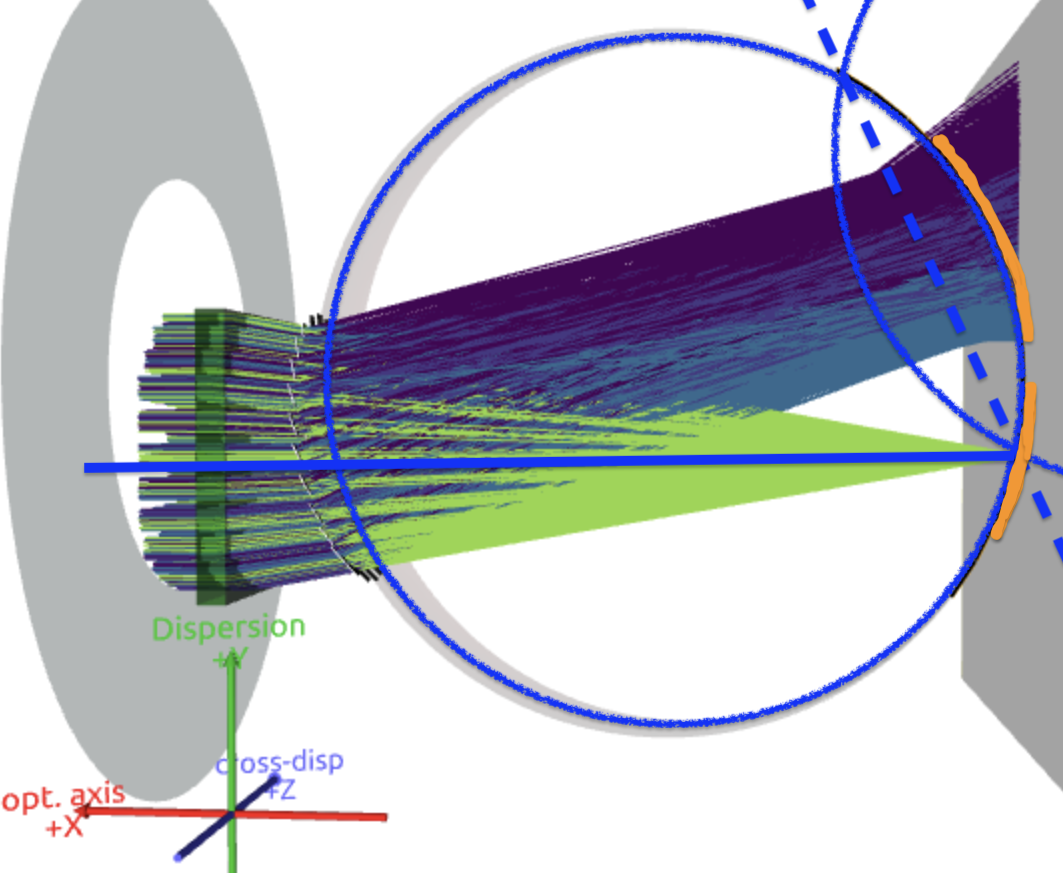}
    \caption{\emph{left:} Sketch of a tilted Rowland torus geometry shown in the plane of the optical axis and the symmetry axis of the torus. Photons arrive from the left. The location of the mirror is shown as blue box. The optical axis (solid blue line) is offset from the center of the Rowland circle (blue square). The symmetry axis of the torus  is marked as blue long-dashed line. The torus is formed by rotating the blue circle around this line. To help visualize the 3D position of the torus blue dotted lines are shown connecting the two circles. $\alpha$ is the angle between the optical axis and the line from the point where the optical axis intersects the Rowland torus (A) to the center of the Rowland circle. $\beta$ is the angle from optical axis to the line connecting A to the ``hinge'' -- the other point in this plane where symmetry axis of the torus and the Rowland circle intersect.
    \emph{right:} Visualization of a ray-trace of the same geometry. Colored lines show the path of individual rays starting on the left. Rays detected in the zeroth order are shown in green, blue and purple are higher orders. Optical axis and the axis of the Rowland torus are indicated in blue and the location of detectors in orange, matching the sketch in the left panel.
        }
    \label{fig:sketch_label}
\end{figure*}

Figure~\ref{fig:sketch_label} shows sketches of the geometry in the plane spanned by the optical axis and the symmetry axis of the torus. Details will be discussed in the following subsections.
\begin{figure*}
    \plotone{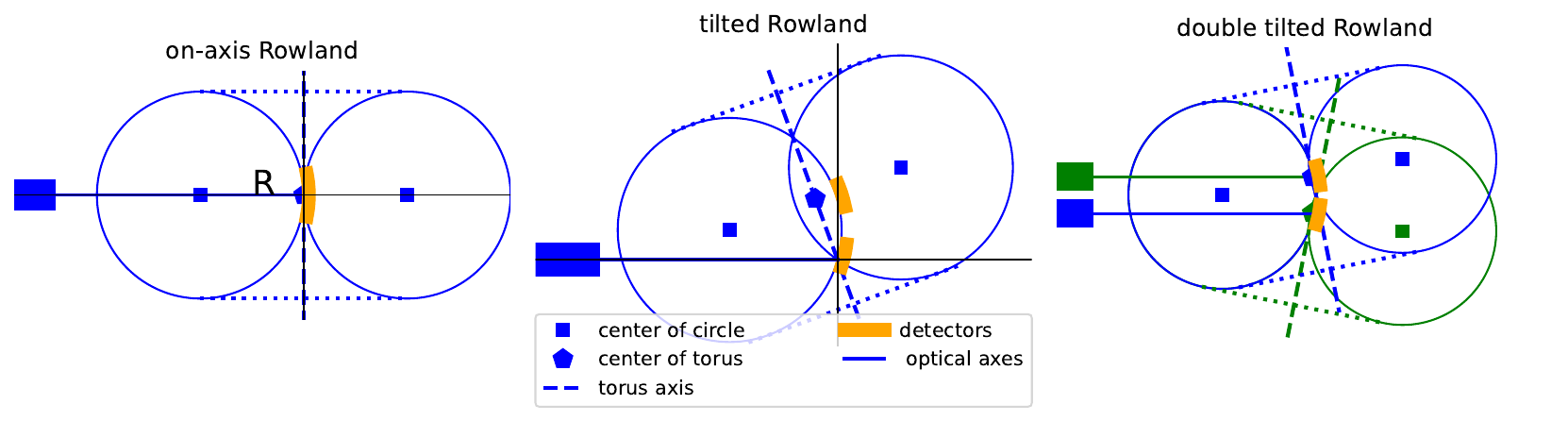}
    \caption{\emph{left:} Sketch of the on-axis Rowland geometry in the plane of the optical axis and the symmetry axis of the torus. Photons arrive from the left.
    \emph{center:} Tilted Rowland torus. The optical axis of the blue channel is offset from the center of the Rowland circle, but intersects the symmetry axis of the torus in the focal point $F$.
    \emph{right:} Layout with two Rowland tori, tilted in opposite direction. In the plane spanned by both optical axes (which are parallel) and the symmetry axis of the tori, the two tori overlap in exactly one circle, the Rowland circle. See figure~\ref{fig:sketch_label} for a more detailed explanation.
        }
    \label{fig:sketch}
\end{figure*}

\subsection{One Rowland Torus with central optical axis}
\label{sect:onetorus}
\begin{figure}
  \begin{interactive}{js}{ifigs/on-axis.zip}
  \includegraphics[width=0.22\textwidth]{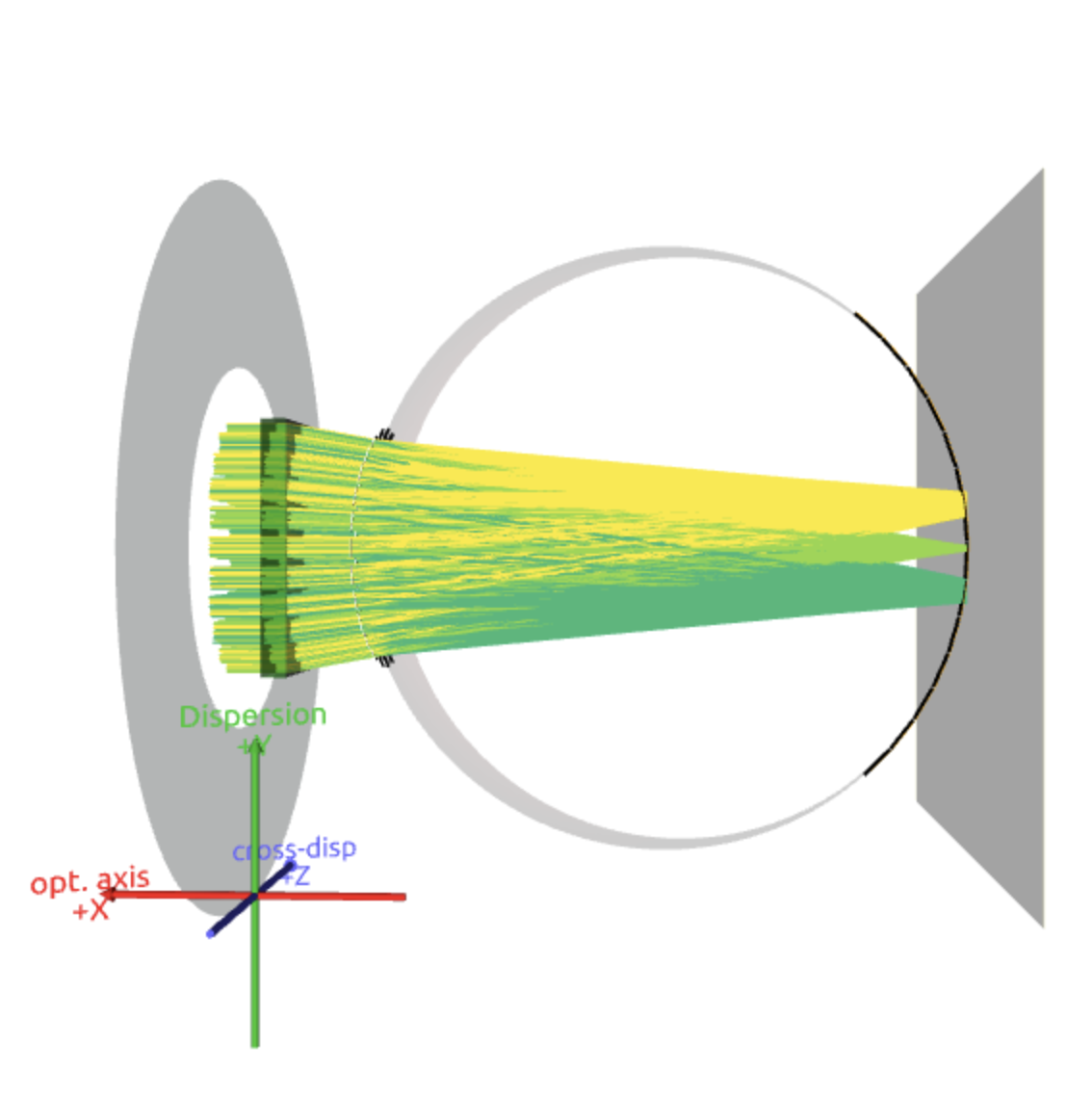}
  \includegraphics[width=0.22\textwidth]{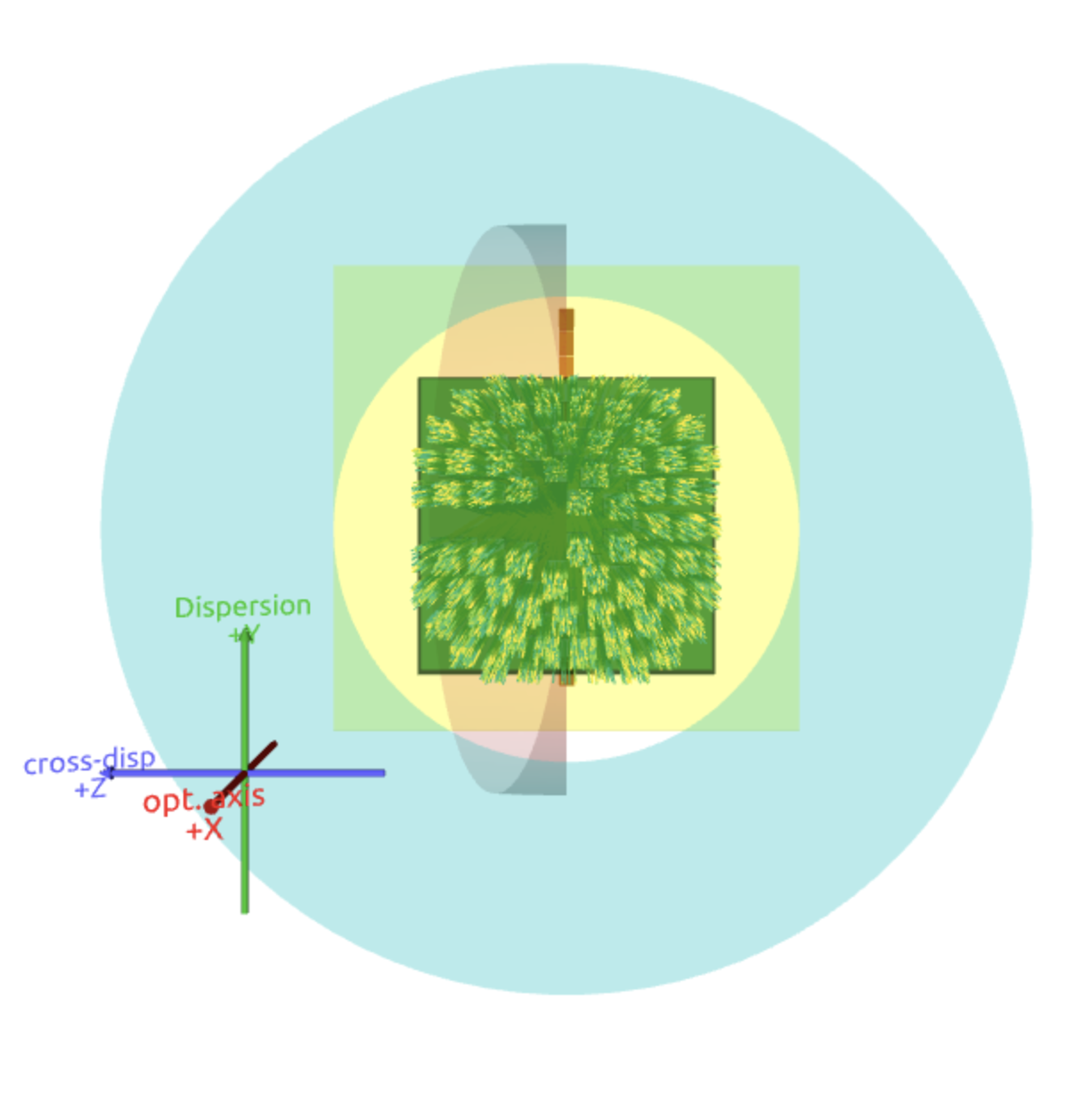}
  \includegraphics[width=0.22\textwidth]{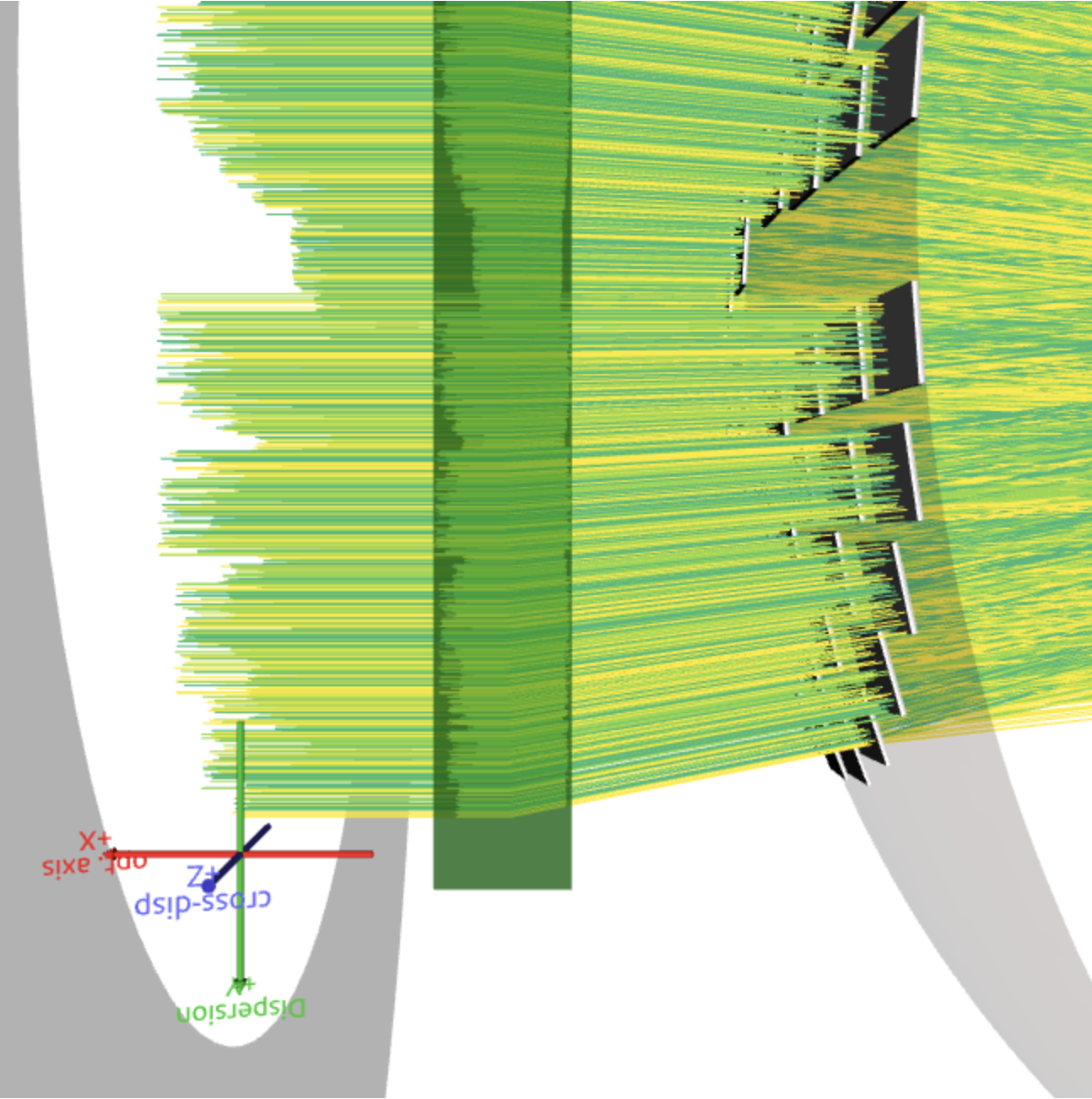}
  \includegraphics[width=0.22\textwidth]{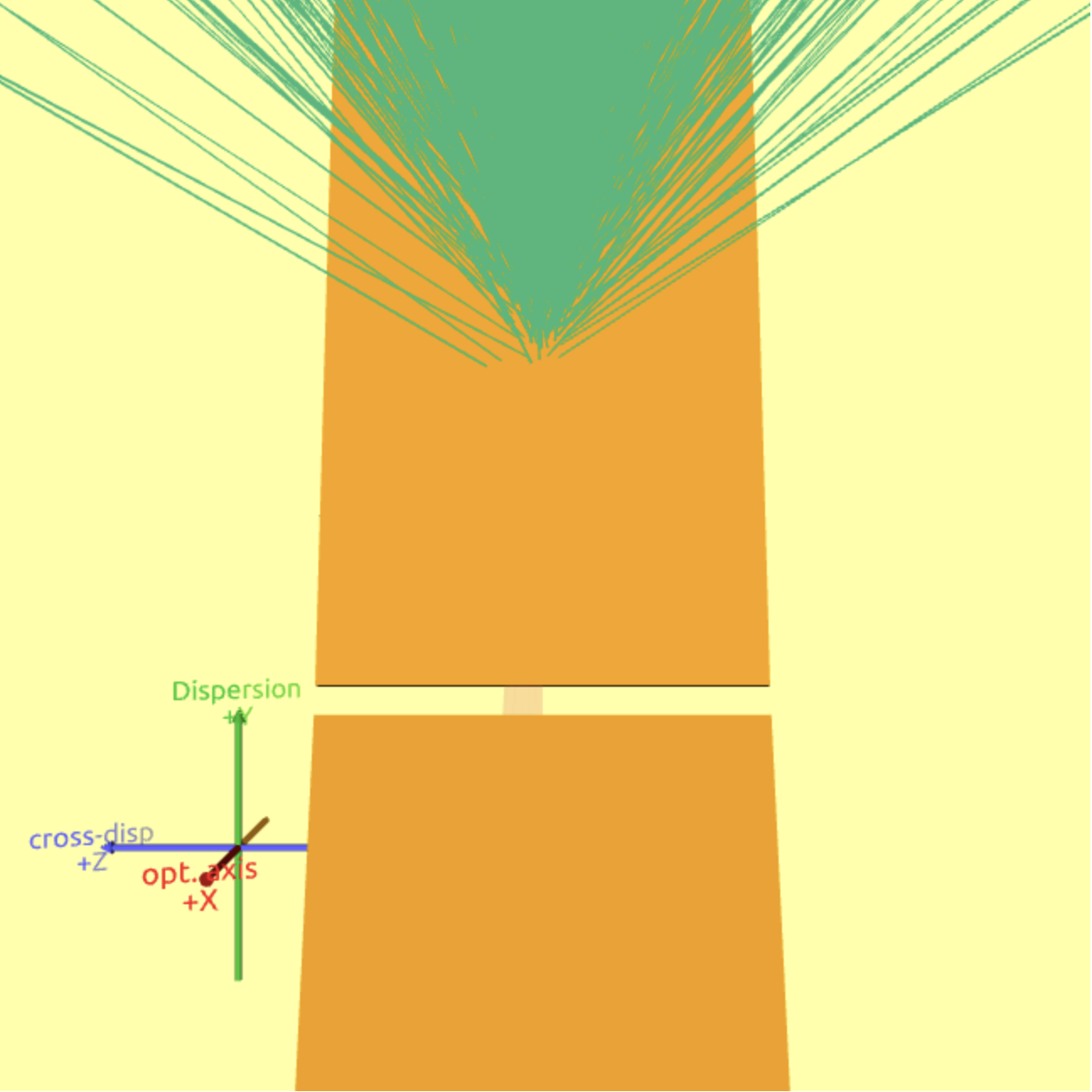}
  \end{interactive}
\caption{3D images of the ray-trace for a source with a continuum spectrum over a limited bandpass, seen (from left to right) from the side, from the top (where the rays enter the aperture), zoom on the grating positions, and zoom on the detector. Rays are shown from the entrance aperture on. They pass a focussing mirror represented by a green box and get diffracted by gratings (white squares). They are detected by CCDs (orange). Rays are colored by the diffraction order: Light green are zero order photons. A section of the Rowland torus is shown as transparent gray. The big square at the right indicates the position of the focal plane. Rays are traced past the detectors to the focal plane to show how the CCDs on the Rowland circle capture the rays where the dispersion width is smallest - past the CCDs the rays widen in dispersion direction again, but still become narrower in cross-dispersion direction. This figure is interactive in the online version, allowing the reader to pan, zoom, and rotate to see those details and inspect the rays from all angles. (Use the mouse to rotate, mouse wheel to zoom, and double-click to focus on a particular element. Right click to navigate to the pre-defined viewpoints shown in the static figure. On other devices, other native controls might be available, e.g.\ two-finger touch to zoom on touchscreens.)
\label{fig:3d:single}}

\end{figure}

For a single Rowland torus with no tilt, the optical axis passes through the center of the Rowland circle and $R=r$ as shown in the left panel of figure~\ref{fig:sketch}. Figure~\ref{fig:3d:single} shows a 3D view of a ray-trace. Since the diffraction gratings are symmetric with respect to the optical axis, positive or negative diffraction orders (up or down in figure~\ref{fig:sketch}) are equally likely.
To avoid the same wavelength to fall in a chip gap on the detector for both the positive and the negative diffraction orders, it is useful to position the CCDs with an offset, so that the chip gaps are not symmetric.

\subsection{One tilted Rowland Torus}


\begin{figure*}
    \centering
    \includegraphics[width=0.9\textwidth]{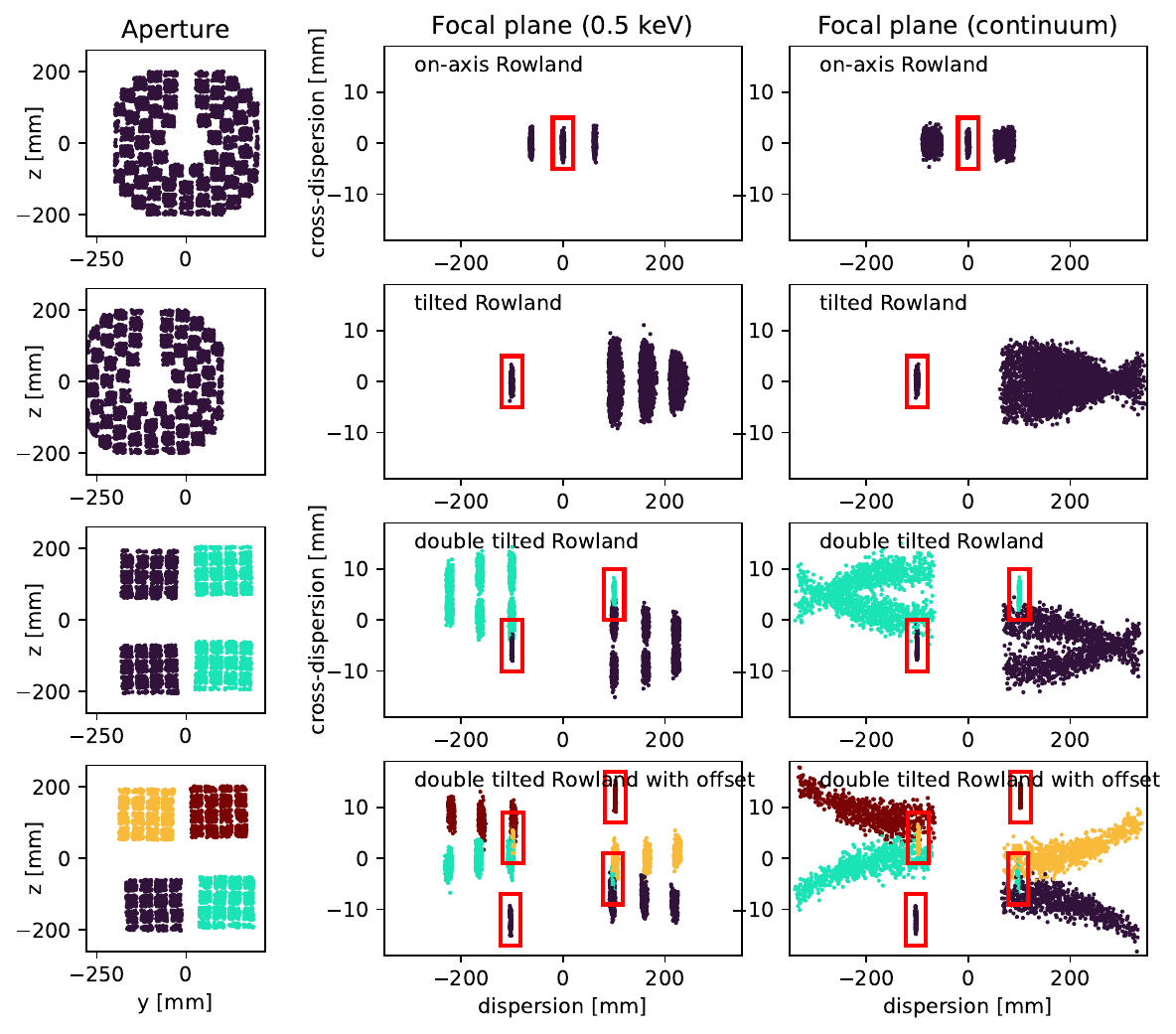}
    \caption{Distribution of photons on the detectors for different designs. \emph{left:} Location where photons pass the aperture and SPOs. \emph{right:} Location photons on a curved detector projected onto the focal plane for a monoenergetic source. Photons are colored by channel and the position of the zero(s) order(s) is highlighted with red boxes. Note that the dispersion (along the $y$ axis) and cross-dispersion ($z$ axis) directions are not to scale. Our toy example detects order -1 and +1 for the on-axis Rowland torus and orders -3, -4, and -5 for the tilted Rowland tori. In the lower two panels the zeroths orders of one channel overlap with the dispersed photons from another channel. This could be avoided with larger offsets between channels in cross-dispersion direction. Also, in real instruments with focal lengths significantly larger than in the illustrative model we use throughout this article, the bending and the width in cross-dispersion direction are typically smaller, so that this is not a problem in practice.
    \emph{right:} Same as in the middle row, but for a continuum source. Again, photons for a specific energy are detected in several orders, but photons from other energies are detected in different locations, so the orders cannot be separated just based on their spatial location.
        }
    \label{fig:fish}
\end{figure*}

\begin{figure}
    \centering
    \includegraphics[width=0.4\textwidth]{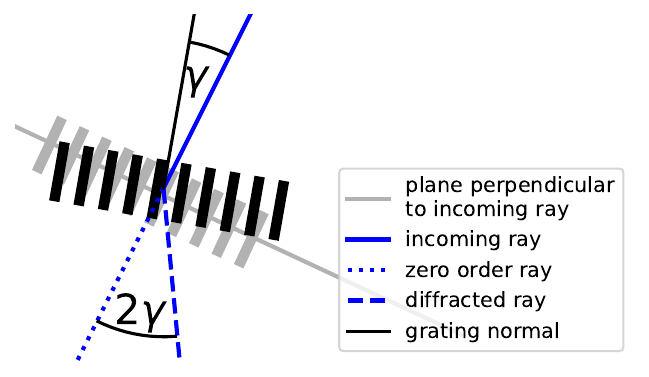}
    \caption{Concept of a blazed grating. The black grating is perpendicular to the incoming rays, the black grating is blazed by and angle $\gamma$ with respect to the incoming rays. The most likely direction for the diffracted photons is near twice the blaze angle.}
    \label{fig:blazed}
\end{figure}

XMM-Newton uses a Rowland torus tilted with respect to the optical axis \citep{2010SSRv..157...15P}; \citet{2010SPIE.7732E..1JH} suggested using the same geometry for Critical Angle Transmission (CAT) gratings. In CAT gratings \citep{2021hai4.book..235H,2022ApJ...934..171H}, the grating bars have a high aspect ratio, that is the distance between two gratings bars is much smaller than the height of the bars. CAT gratings are used at a blaze angle, where the grating bars are inclined with respect to the incoming rays (see figure~\ref{fig:blazed}). CAT gratings preferentially diffract photons to twice the blaze angle and can reach high diffraction efficiencies: \citet{2022ApJ...934..171H} measured $>20$\% summed over orders 4-7 at 1.8~keV. However, the blaze angle also means that the gratings can deviate significantly from the surface of the  Rowland torus. To minimize this deviation, the Rowland torus can be tilted.
This tilt can be described by two angles $\alpha$ and $\beta$ as shown in figure~\ref{fig:sketch_label}. $\alpha$ is the angle between the optical axis and the line from the point where the optical axis intersects the Rowland torus (labelled ``A'' in the figure) and the center of the Rowland circle.  We choose the distance between point A and the focal point F to be slightly shorter than the focal length $f$ of the mirror to allow for space to mount the gratings behind the mirror and use the symbol ``a'' for the distance between A and F.
In this configuration, the torus has $R < r$ so there are two points where the torus overlaps itself in the plane spanned by the optical axis and the symmetry axis of the torus (the plane shown in figure~\ref{fig:sketch_label}). The first of those intersection points is F. We call the second point ``Hinge''.
$\beta$ is the angle from optical axis to the line connecting A and the Hinge. From this we can derive the radius of the Rowland circle as
$$r = \frac{a}{2\cos \alpha}.$$
 If the focal point F is the origin of the coordinate system, then the center of the Rowland circle (blue square in left blue circle in figure~\ref{fig:sketch_label} and \ref{fig:sketch}) is at location
 \begin{equation}
    \begin{pmatrix} x \\ y \\ z \end{pmatrix} =
    \begin{pmatrix} \frac{a}{2} \\ \frac{a}{2}  \tan\alpha \\ 0\end{pmatrix}.
 \end{equation}

(In this coordinate system the astrophysical source is located at $x=+\infty$.)
Since A, F, and the Hinge are by definition all on the Rowland circle, their distance to this point is $r$. With this symmetry, we can obtain the angle of the axis of the torus with respect to the $y$-axis. This angle is $\beta-\alpha$. With that, we can derive $R$, the distance between the center of the Rowland circle and the center of the torus, to
\begin{equation}
    R = r \sin\left(\frac{\pi}{2} - \beta\right) = r \cos\beta\ ,
\end{equation}
which places the center of the torus at
\begin{equation}
    \begin{pmatrix} x \\ y \\ z \end{pmatrix} =
    \begin{pmatrix}
        \frac{a}{2}-R\cos{(\beta-\alpha)} \\
        \frac{a}{2}\tan{\alpha}+R\sin{(\beta-\alpha)}
        \\ 0
        \end{pmatrix}.
\end{equation}

\begin{figure}
    \begin{interactive}{js}{ifigs/tilted.zip}
    \includegraphics[width=0.22\textwidth]{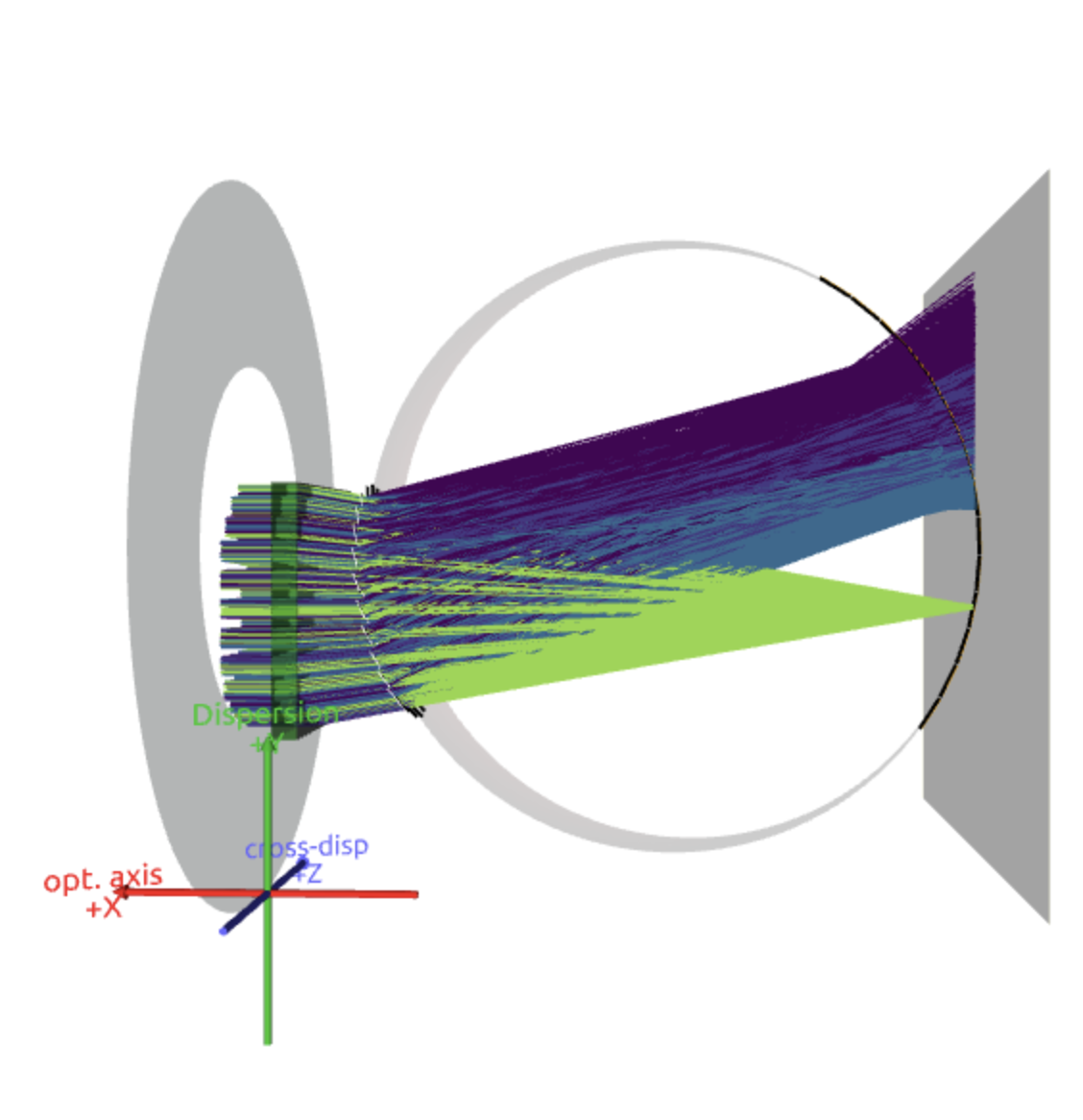}
    \includegraphics[width=0.22\textwidth]{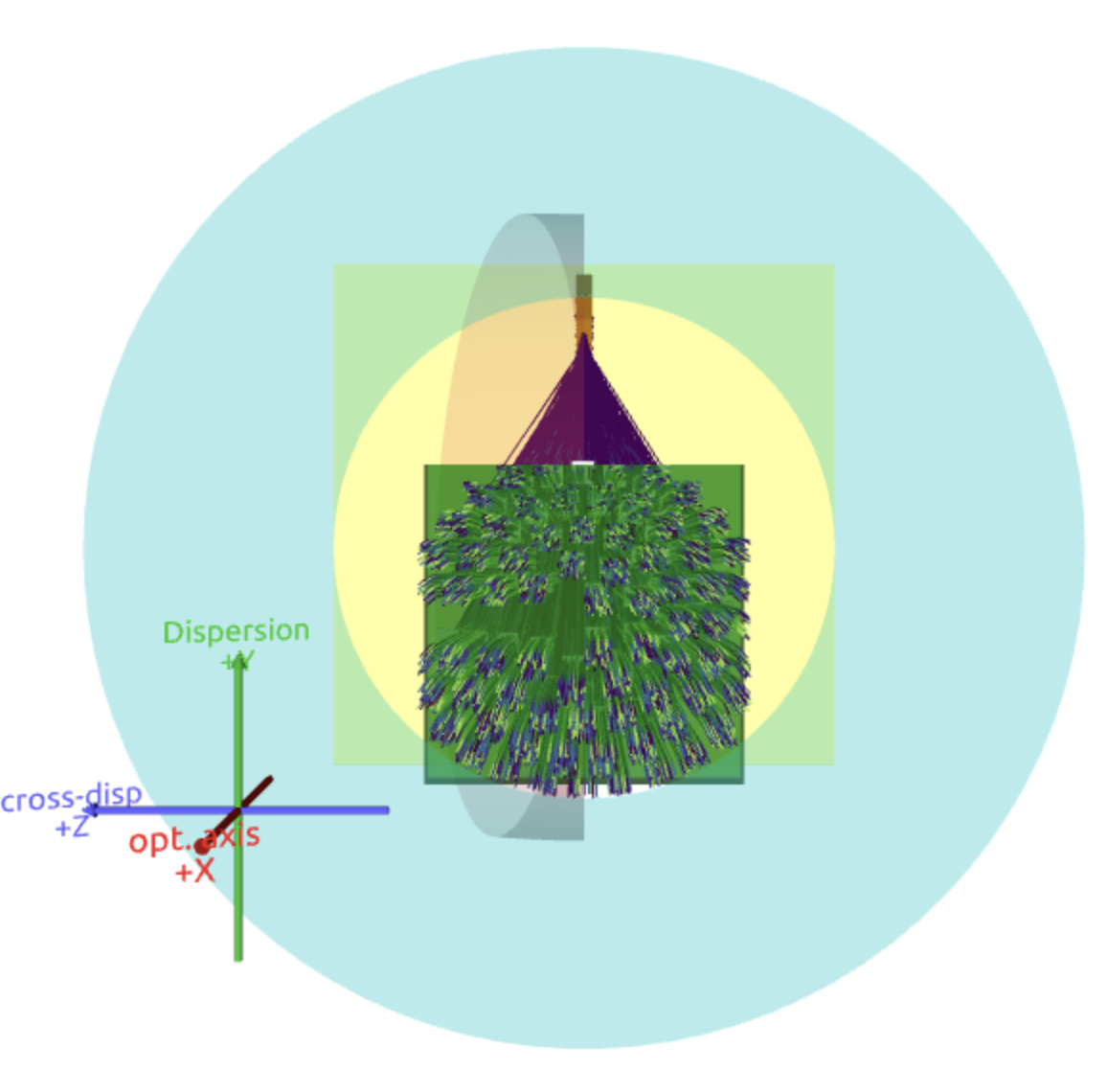}
    \includegraphics[width=0.22\textwidth]{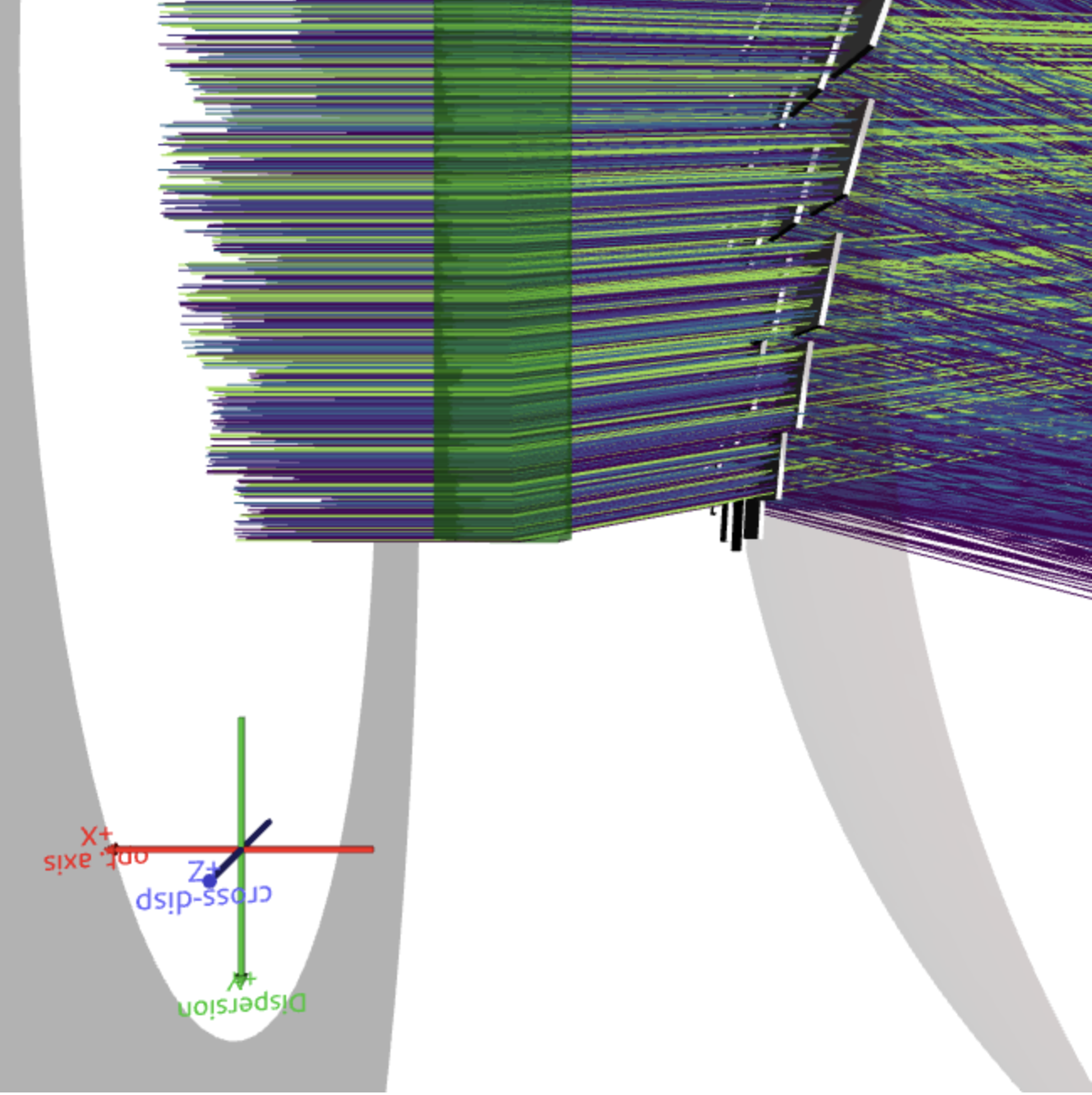}
    \includegraphics[width=0.22\textwidth]{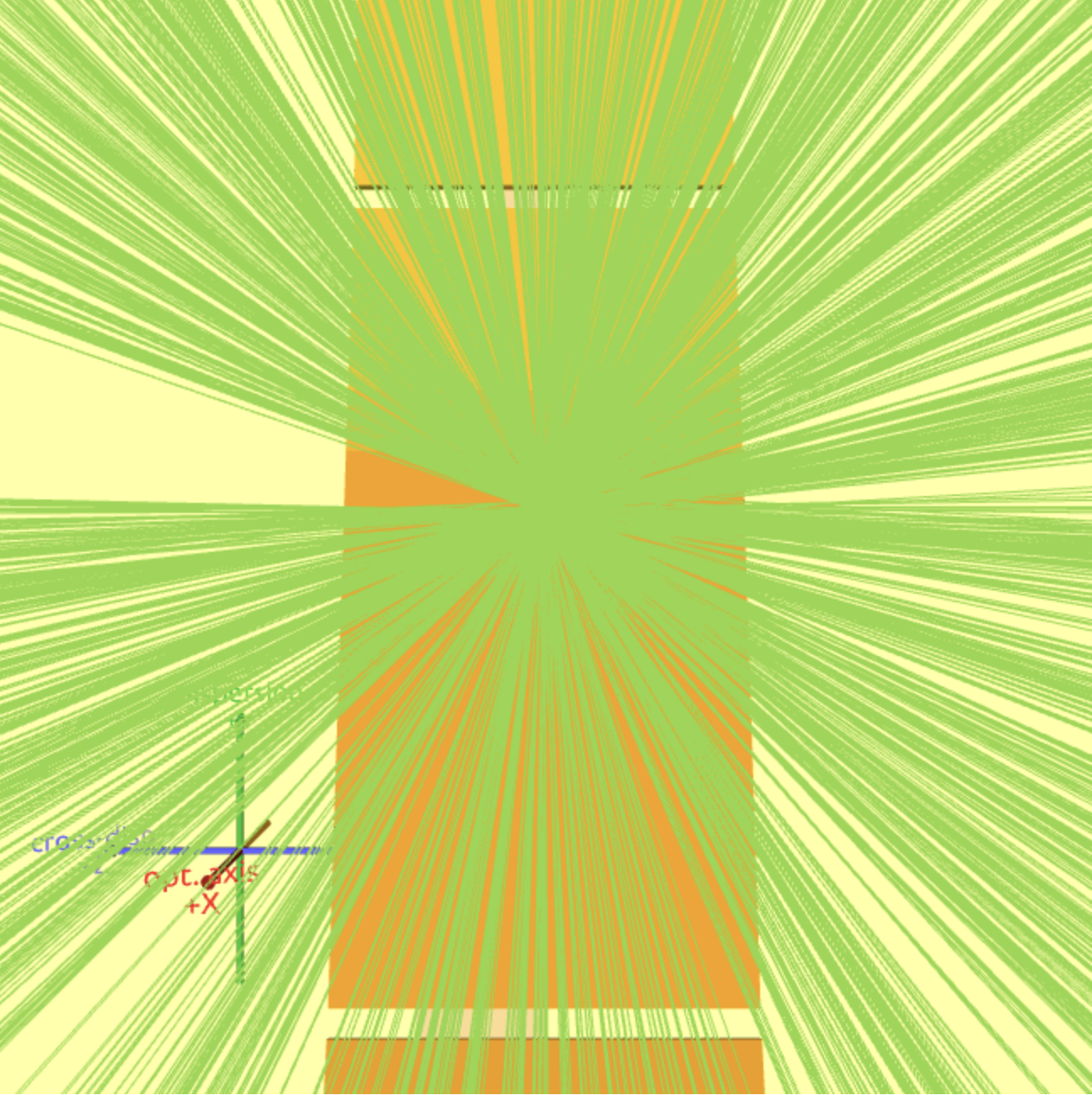}
    \end{interactive}
  \caption{Ray-trace like figure~\ref{fig:3d:single}, but for a tilted Rowland torus setup. See figure~\ref{fig:3d:single} for an explanation of the elements shown in the image and the interactivity in the online version.
  }
  \label{fig:3d:tilted}
  \end{figure}

For real CAT gratings typical blaze angles are 1-2~deg. Figure~\ref{fig:3d:tilted} shows this layout in 3D, but with a very large blaze angle for clarity. Tilting the Rowland torus by about twice the blaze angle reduces the average deviation of the grating position from the surface of the Rowland torus.
The Rowland geometry optimizes the spectral resolving power by minimizing the spread of the photons in dispersion direction; detectors are located on the Rowland circle, which represents the position of minimal spread in dispersion direction. This is not the same location as the imaging focal plane, which is defined by minimizing the total size of the 2D PSF. At the focal point, the Rowland circle and the imaging focal plane match. Figure~\ref{fig:fish} (right column) shows how the dispersed photons spread out in cross-dispersion direction with increasing distance from F, roughly resembling the outline of a fish. With a tilted Rowland torus, there is a second intersection point between the spectral and the imaging focus where the cross-dispersion width becomes small (the point between the body and the tail of the fish, the ``Hinge'', about +300~mm in dispersion direction in the second panel of figure~\ref{fig:fish}). This location can be chosen to be roughly in the center of the dispersed spectrum and thus limit the cross-dispersion profile to reduce the area of the detector that contributes to the extracted signal and thus the background. We note that the width in cross-dispersion direction is much larger for larger diffraction angles and, in real instruments, will be smaller than shown in the figure.

A tilted Rowland torus can also be combined with sub-aperturing. 
In this case, not all the fish shape is filled with photons. This can be seen by comparing the dark purple points in the second, third, and fourth panel in figure~\ref{fig:fish}, right column. The Rowland torus for this channel has the same tilt and dimensions in all three designs, but in the second panel, the mirror fills the full circle, in the third panel it has just two rectangles, and in the fourth panel just one. In either case, the spectral extraction region used in data analysis should take this shape into account to minimize detector background.

\subsection{Double tilted Rowland Torus}

Sub-aperturing reduces the geometric area of the mirror and thus the effective area of the instrument. This can be compensated by adding a second channel, with mirrors mounted next to the mirrors of the first channel. Again, gratings and detectors for this second channel need to be positioned on a Rowland torus. However the two Rowland tori for the two channels need not be identical. When $R> r$ the Rowland circle intersects the focal plane twice. We choose those two points as the focal points for the two channels of the instrument; this layout is symmetric with respect to the center of the Rowland circle (figure~\ref{fig:sketch}, right). By selecting $\alpha$ close to the average dispersion angle of dispersed photons, we can reach a configuration where the same detectors that detect the zeroth order of one channel, will now also detect the dispersed photons of the other channel and vice versa. The second channel also adds redundancy in case of detector failure, and can compensate for chip gaps if the position of the detector elements on one channels is moved along the Rowland circle a little bit with respect to the other channel. 

\subsection{Channel offsets}
\begin{figure}
    \begin{interactive}{js}{ifigs/offset.zip}
    \includegraphics[width=0.22\textwidth]{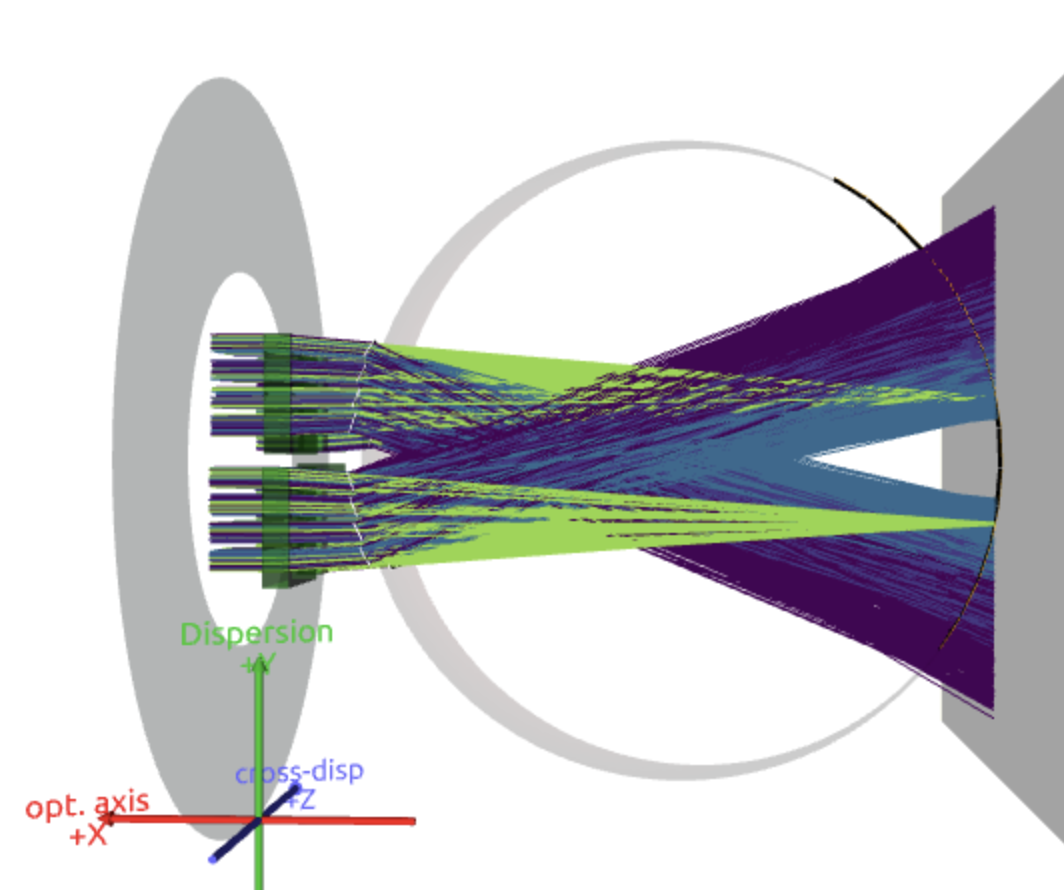}
    \includegraphics[width=0.22\textwidth]{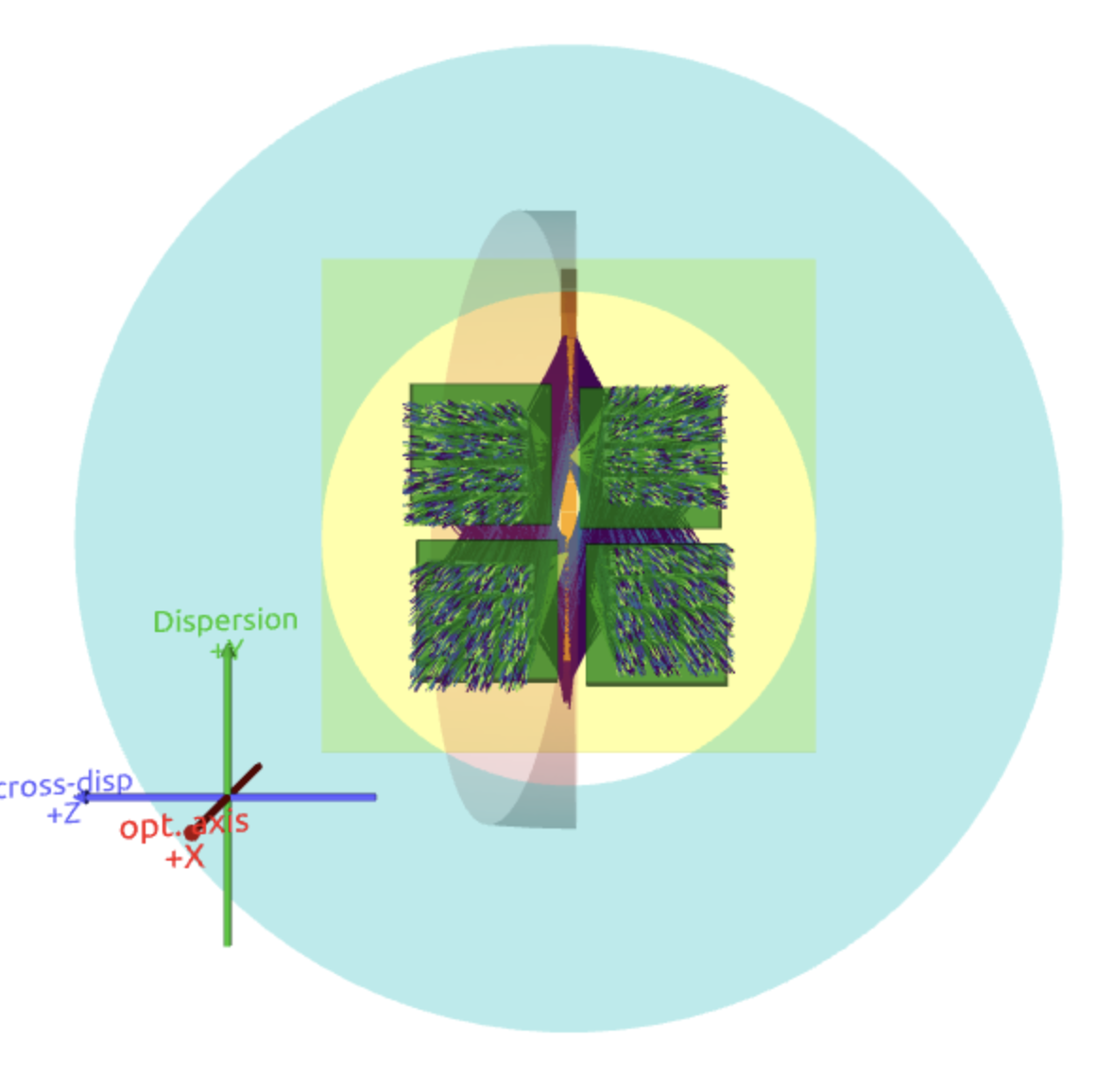}
    \includegraphics[width=0.22\textwidth]{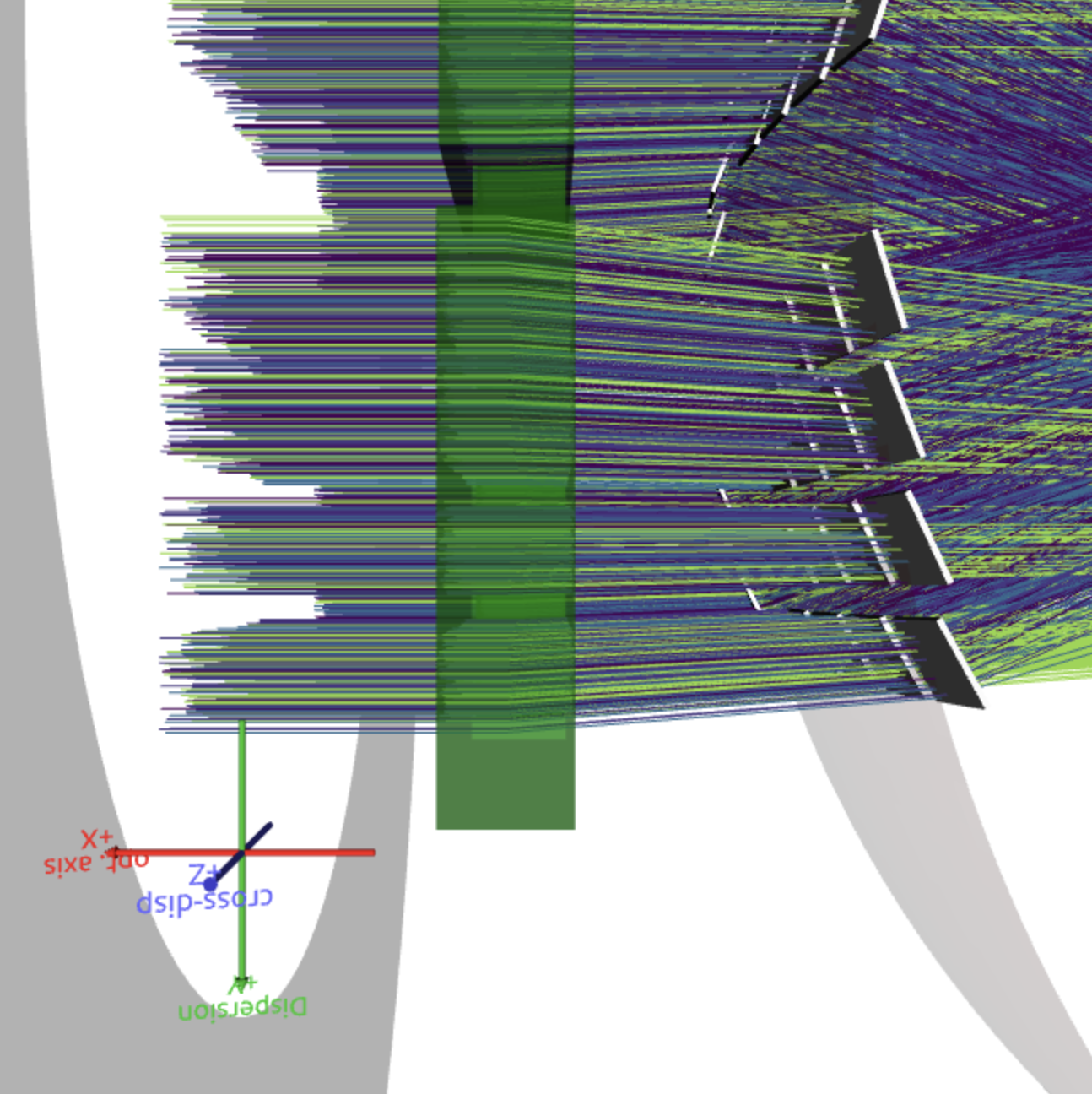}
    \includegraphics[width=0.22\textwidth]{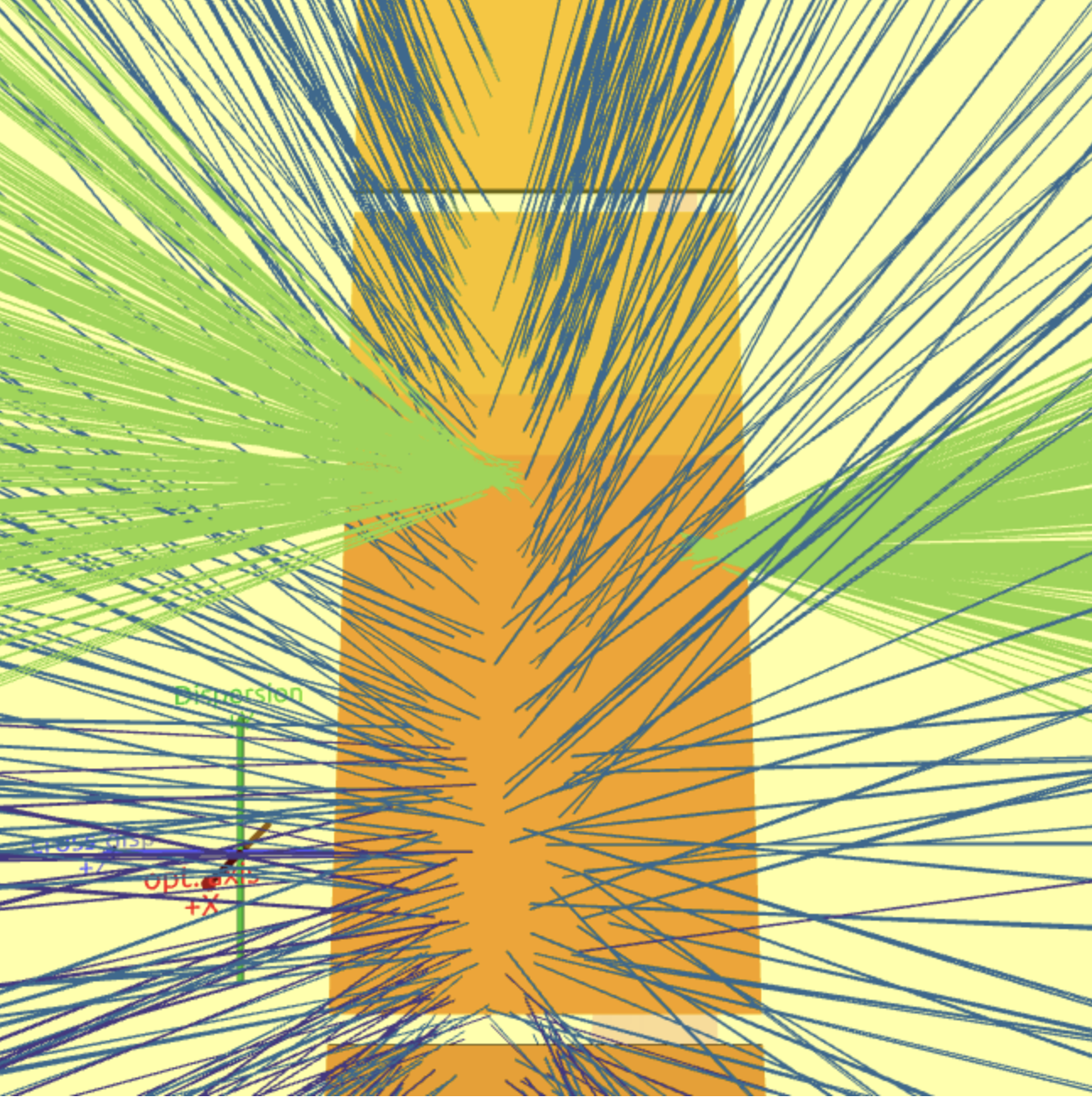}
    \end{interactive}
    \caption{Ray-trace like figure~\ref{fig:3d:tilted_double}, but with offsets in dispersion and cross-dispersion direction. See figure~\ref{fig:3d:single} for an explanation of the elements shown in the image and the interactivity in the online version.
        }
    \label{fig:3d:tilted_double_offset}
\end{figure}

The next two subsections explain how the channels can be offset in both dispersion and cross-dispersion direction to reduce alignment requirements and mitigate chip gaps. Figure~\ref{fig:3d:tilted_double_offset} shows a 3D ray-trace image in this configuration; the resulting signal on the detector is shown in figure~\ref{fig:fish} (bottom row).

\subsubsection{Channel offsets in the cross-dispersion direction}
In the data reduction, we need to assign a unique wavelength to each dispersed photon. To do that, it is necessary to know which optical channel a photon passed through and thus the setup needs to avoid overlapping photons from both channels on the same location of the detector. This can be accomplished by shifting the Rowland tori (and mirrors and gratings positioned on them) in the cross-dispersion direction (into or out of the plane of the plot in figures~\ref{fig:sketch_label} and~\ref{fig:sketch}, i.e.\ the $z$ direction of our coordinate system). The shift should be large enough to clearly separate the two channels, but keep both dispersed spectral traces on the detector with margin (figure~\ref{fig:fish}). Since this shift is perpendicular to the Rowland circle, it does not impact the spectral resolving power.

The DTRS layout is particularly well-suited for sub-aperturing. In this case, each channel will have two opposing rectangles filled with mirror elements and gratings leading to a total of four mirror and grating rectangles. These can then be arranged closely together (see figure~\ref{fig:fish}) to maximize the area used given the size constraints of the spacecraft. In this case, it is an option to split each of the two channels into two sub-channels with their own separate offset, such that there are two parallel traces running in each direction (Figure~\ref{fig:3d:tilted_double_offset}) for a total of four traces (figure~\ref{fig:fish}, bottom panel). Without this split, the opposing mirror and grating wedges have to be carefully aligned to the same optical axis, and any misalignment between them would broaden the PSF and thus reduce the spectral resolving power. With this split, each wedge defines its own optical axis and the exact location, and thus the wavelength scale for each zeroth order can be determined from observations.

\subsubsection{Channel offsets in the dispersion direction}
When each channel is split into two sub-channels as described above and each of the sub-channels has its own Rowland torus, 
the two sub-channels can also be offset in the dispersion direction. This causes the Rowland torus that defines the grating positions to deviate from the torus that sets the position of the detectors, but if the offset is small, the impact on the spectral resolving power is negligible. A benefit of such an offset in dispersion direction is that the dispersed spectra of the two sub-channels do not hit chip gaps at the same wavelength. 

\section{Discussion}
\label{sect:discussion}

\subsection{Advantages of the DTRT concept}
\subsubsection{Subaperturing space considerations, and resolving power}
Manufacturing mirrors with a smaller PSF is more difficult and costly than manufacturing mirrors with a larger PSF but otherwise identical properties. For a spectrograph, the width of the PSF in dispersion direction is far more important than the width in cross-dispersion direction. If the width in dispersion direction can be reduced by sub-aperturing, a higher spectral resolving power is achieved at lower cost and technical complexity compared to reducing the PSF of the mirror itself. On the other hand, it reduces the collecting area of the mirror, since only a fraction of the full circle can be used. The DTRS design allows us to compensate for this by having multiple channels, ultimately delivering better resolving power at the same mirror requirements than a single-channel design. The independent mirrors in a DTRS design can be arranged on the front-end of the spacecraft such that they take up a similar geometric area as a mirror filling the full circle in a single channel design.

An alternative design is to use a single, circular mirror and sub-aperture by dispersing light from different parts of the mirror to different, independent detectors rotated in the focal plane \citep{2019JATIS...5b1003G}.

\subsubsection{Zeroth order is seen}
Observing the direct image of all channels on the detector has several advantages. First, most of the high-energy photons are neither diffracted nor absorbed by typical CAT gratings. So, the direct image can be used for analysis of high-energy spectra using the intrinsic energy resolution of the detector, e.g.\ a CCD or even microcalorimeter. For example, the Fe line at 6.7~keV is a widely used diagnostic for many astrophysical objects; in a DTRT design this line can be observed with an energy sensitive detector such as a CCD in the zeroth order, while the lower-energy photons are observed in the dispersed orders at the same time. Second, accurate knowledge of the position of the direct image is important for the wavelength calibration of the dispersed signal. The wavelength of the light is determined by measuring the distance $x$ from the detected photon to the zeroth order and the known distance $a$ between the gratings and the focal plane. Conceptually, $\tan \theta \approx x/a$. With the known grating constant $d$, equation~\ref{eqn:diffraction} then yields $\lambda$. For high spectral resolving power and large values of $\theta$ it might be important to take the 3d shape of the detector into account, but either way, accurate knowledge of the zero order position is crucial.

\subsubsection{Robustness}
The DTRS concept increases the robustness and reduces the technical risk of a mission in several ways. Because we can assign each sub-apertured mirror wedge its own channel (in the 4 channel design with offsets), there is no need to align the mirrors for each channel to the other channel to better than a cm or so.
Also, the design is robust against loosing individual detectors in a camera. Since half of the photons are dispersed from left-to-right and the others right-to-left, loss of a detector element in a camera does not result in a complete loss of spectral coverage for any wavelength. The multichannel design can also help to uncover systematics in the calibration.


\subsection{Drawbacks}
\subsubsection{Number of detector elements}
An instrument that uses only a single tilted Rowland torus needs to cover a long strip of dispersed light with a detector and a small area around the zeroth order. On the other hand, a DTRS needs two long detector strips to cover the dispersed signal on both sides; those cameras will cover the zeroth order of all channels at the same time.

\subsubsection{Background and source confusion}
Because the DTRS concept spreads the dispersed photons over a larger detector area than a single-channel design, the background contamination in the spectra is higher, simply because more pixels contribute to the extracted science spectrum. This does not matter for a bright isolated point-source, but makes observations of faint sources, where the dispersed spectrum has count rates comparable to the background, difficult. Similarly, observations are difficult when extended emission or many point sources are in the field-of-view, for example in the galactic center or in star forming regions. In this case, spectra from different locations on the sky overlap on the detector.

\subsubsection{Complexity of spectral fitting}
Depending on the detector technology chosen, the DTRS concept might also lead to undesirable coupling between the spectra of different channels. For example, traditional CCDs have a read-out streak, where photons that arrive at the detector during the charge transfer are not associated with the correct position on the detector in the direction of the read-out. This means that photons that should be recorded at the zero-order position or at the position of a strong emission line in one channel are assigned to the wrong channel, where they might cause what looks light a weak emission feature. This cross-talk can be avoided by using different detector technologies, or be taken into account in the data analysis. Similarly, the pure number of channels with different chip gaps and detector positions increases the computational complexity of the data analysis compared to a single-channel design. Development of dedicated software tools is needed to help with the analysis of these multichannel data.

\subsection{Similar concepts}

The Far Ultraviolett Spectroscopic Explorer (FUSE) \citep{2000ApJ...538L...1M} faced a design challenge similar to what we discuss here for X-ray instruments. FUSE also has four spectroscopic channels with gratings arranged in a Rowland geometry. Unlike in X-rays, UV gratings can operate in reflection close to normal incidence. The reflection gratings are thus positioned almost perpendicular to the path of the rays tangential to the surface of the Rowland torus. Thus, a few gratings (just one per channel in the extreme case) are sufficient and, and even if the grating itself is flat, it never deviates much from the surface of the Rowland torus. Since UV light has a longer wavelength, it can be dispersed to larger angles increasing the spectral resolving power. In FUSE, detectors just cover the dispersed signal, and not the direct beam (zeroth order) and detectors for the different channels are independent of each other.

Our design of a DTRS places the direct beam for all channels on a detector. This way, the position of the pointing can be constantly monitored and determined from the observed data. If the pointing position drifts and the optical channels are not fully aligned, this can be seen in the science data. In contrast, FUSE has performed observations where the target was not in the field-of-view of all channels, but the amount of the exposure time lost is unclear \citep{2000ApJ...538L...1M}.

\section{Applications}
\label{sect:applications}
As a concrete example, we describe the Arcus X-ray spectrograph  \citep{2023SPIE12678E..0ES}. Arcus has gone through multiple iterations for different mission calls, most recently as a NASA X-ray Probe concept submitted in 2023. This mission adopts a DTRS design with $\alpha=3.6$~deg and $\beta=2\alpha$. It uses silicon-pore optics (SPO) with a 12~m focal length that are developed for the Athena mission \citep{2023SPIE12679E..05G}; $a$ is slightly smaller than the focal length with $a=11876$~mm. SPOs are manufactured as small stacks of about 30 mirror plates with sizes of order 10-20~cm in each direction (the exact size depends on the radius). 40 SPOs are aligned into one petal and each petal defines one of four optical channels in a DTRS design with offsets. The two sub-channels are offset by $\pm2.5$~mm in dispersion direction with respect to the design with just two Rowland tori. In cross-dispersion the four channels are located at $-7.5$, $-2.5$, $+2.5$, and $+7.5$~mm from the center of the CCDs.
CAT gratings are manufactured through deep reactive-ion etching of silicon-on-insulator wafers \citep{2022ApJ...934..171H,2023SPIE12679E..0LH}. These gratings are mounted with a blaze angle, i.e.\ they are intentionally tilted by 1.8~deg with respect to the incoming photons, which concentrates the diffracted photons around the ``blaze peak'' at a diffraction angle of $2 * 1.8 = 3.6$~deg, so that only 8~CCDs are needed in each camera covering about 30~cm in the dispersion direction with a gap close to 50~cm between the two cameras. Arcus will achieve a spectral resolving power around 3500 for wavelengths between 1.5 and 6.0~nm, averaged over all diffraction orders, and an effective area up to 500~cm$^2$ between 1.5 and 2.5~nm, dropping to 100~cm$^2$ at 6.0~nm. Detailed ray-tracing simulations for Arcus that optimize the parameters of the design and predict the performance depending on the misalignment for the different components are shown in \citet{2017SPIE10397E..0PG,2018SPIE10699E..6FG,2023SPIE12678E..1DG}.

\begin{figure*}
    \plotone{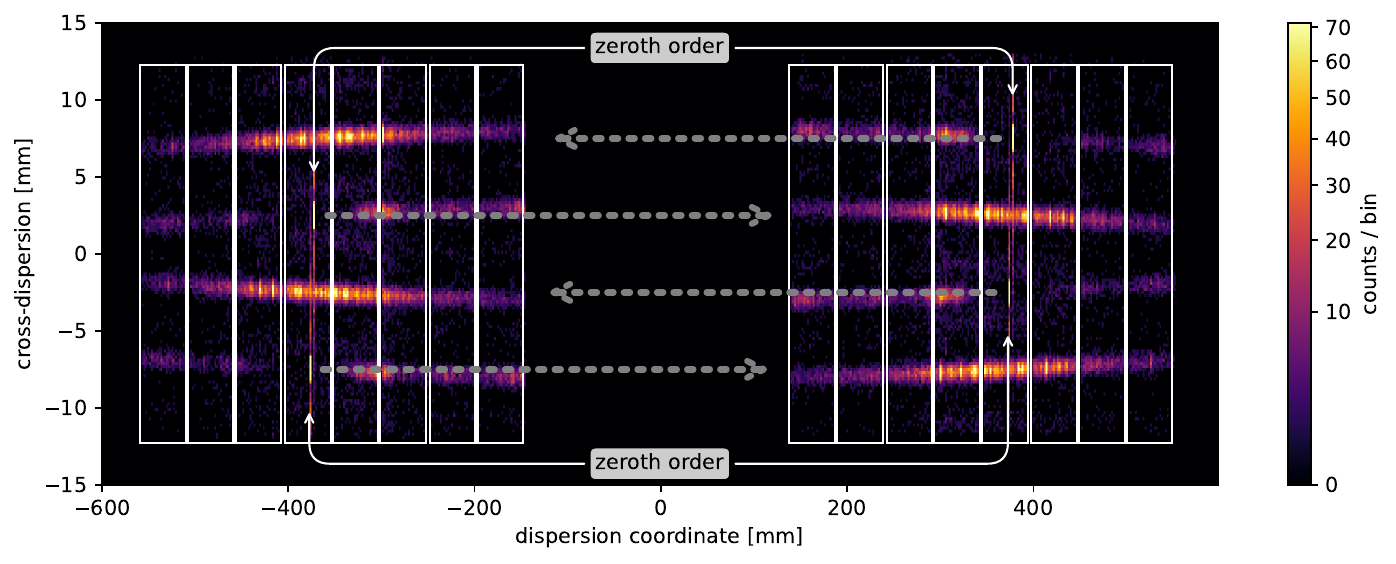}
    \caption{Simulation of an Arcus observation of an emission line spectrum (see section~\ref{sect:applications} for details). Positions of the CCDs are outlined with white frames and the zeros orders are indicated. Dashed arrows mark the dispersion direction of the four spectra from the zeroths order. Only photons that are detected on a CCD are included. Note that the dispersion and cross-dispersion directions are scaled differently. The cross-dispersion direction is parallel to the coordinate system $z$ axis. The dispersion direction is measured along the curved Rowland circle; at any given location on this plot is within 5~deg of the coordinate system $y$ axis.
        }
    \label{fig:Arcusfull}
\end{figure*}

Figure~\ref{fig:Arcusfull} shows a simulation of an Arcus observation of an emission line spectrum. The spectral model is taken from the active star EQ~Peg~A \citep{2008A&A...491..859L} and the simulation is run for an exposure time of 100~ks. Note that the plot shows an extremely distorted view of the CCD plane. The dispersion axis extends over 1.2~m, while the cross-dispersion axis covers only 2.5 cm.

This particular simulation does not include any background. The photons that are seen in between the spectra are cross-dispersed photons. The CAT gratings disperse photons in one direction. However, the main CAT grating bars are held in place by a support structure (L1) that also acts as a grating and disperses some photons perpendicular to the main dispersion direction. This is visible in the plot for the zeroth orders and the strongest emission lines.




\section{Summary}
\label{sect:summary}
\emph{Chandra}/HETGS and LETGS employ an on-axis Rowland torus geometry. We show how this geometry can be modified in several steps to increase the spectral resolving power without compromising the effective area for a specific class of future X-ray observatories. Our design is optimized for grating spectroscopy of X-ray point sources. We use sub-aperturing to increase the spectral resolving power. We tilt the Rowland torus such that the optical axis of the mirror is offset from the center of the Rowland circle. That reduces the deviation of flat, tilted gratings from the surface of the torus and also allows us to position a second channel next to the first one, doubling the effective area compared to a single, sub-apertured channel. Both channels have different Rowland tori (double tilted Rowland spectrograph, DTRS), but they overlap in the Rowland circle, and thus both channels can be imaged onto the same set of detectors. In this geometry, all zeroths orders are visible for wavelength calibration and to detect hard X-rays in direct imaging. Finally, we discuss how small offsets between the channels can mitigate chip gaps and reduce the alignment requirements between optical elements during the assembly of the instrument. We illustrate these concepts with sketches and ray-traces.

We briefly discuss Arcus, one NASA Probe class mission concept that implements a DTRS and show realistic ray-traces for a 100~ks observation of an emission line dominated spectrum observed by Arcus.

\begin{acknowledgements}
Support for this work was provided in part through NASA grant 80NSSC22K1904 and Smithsonian Astrophysical Observatory (SAO)
contract SV3-73016 to MIT for support of the {\em Chandra} X-Ray Center (CXC),
which is operated by SAO for and on behalf of NASA under contract NAS8-03060.
\end{acknowledgements}

\software{marxs \citep{2017AJ....154..243G}, AstroPy \citep{2013A&A...558A..33A,2018AJ....156..123A}, NumPy \citep{harris2020array}, Matplotlib \citep{Hunter:2007}, IPython \citep{IPython}}

\bibliography{bib}{}
\bibliographystyle{aasjournal}


\end{document}